\documentstyle[prd,aps]{revtex} 

\input psfig \pssilent
\def\wsymbol#1/#2/#3/#4/#5/#6/{
     \left\{  \matrix{#1 & #2 & #3 \cr #4 & #5 & #6 \cr}  \right\}}
\begin{document}
\title{A finite  spin-foam-based theory of three and four dimensional 
quantum gravity}
\author{Rodolfo Gambini$^1$, and Jorge Pullin$^2$}
\address{1. Instituto de F\'{\i}sica, Facultad de Ciencias, 
Universidad
de la Rep\'ublica, Igu\'a esq. Mataojo, Montevideo, Uruguay}
\address{2. Department of Physics and Astronomy, 
Louisiana State University, 202 Nicholson Hall, Baton Rouge,
LA 70803}

\date{February 11th 2002}
\maketitle
\begin{abstract}
Starting from Ooguri's construction for $BF$ theory in three (and
four) dimensions, we show how to construct a well defined theory with
an infinite number of degrees of freedom.  The
spin network states that are kept invariant by the evolution operators
of the theory are exact solutions of the Hamiltonian constraint of
quantum gravity proposed by Thiemann. The resulting theory is the
first example of a well defined, finite, consistent, spin-foam based
theory in a situation with an infinite number of degrees of
freedom. Since it solves the quantum constraints of general relativity
it is also a candidate for a theory of quantum gravity. 
It is likely, however, that the solutions constructed correspond
to a spurious sector of solutions of the constraints. The richness of
the resulting theory makes it an interesting example to be analyzed by
forthcoming techniques that construct the semi-classical limit of spin
network quantum gravity.
\end{abstract}

\section{Introduction}

Attempts to construct a well defined and consistent theory of quantum
gravity have recently received a significant boost through the
introduction by Ashtekar and Lewandowski \cite{AsLe} of mathematical
tools for performing well defined calculations in the context of
theories of connections modulo gauge transformations in infinite
dimensional situations. The introduction of these mathematical tools
has actually had impact in two different avenues of quantization of
general relativity: the canonical and the covariant (path integral)
approach.

In the canonical approach, Thiemann \cite{Th} was able to construct a
finite, well defined, anomaly-free representation of the quantum
Hamiltonian constraint. In a separate development, similar techniques
were used to define an equally consistent operator on the space of
Vassiliev knot invariants \cite{DiGaGrPu}. Thiemann's Hamiltonian
operates on a space of diffeomorphism invariant spin networks. The
algebra of two Hamiltonians with different lapses is therefore an
Abelian one, and it is faithfully implemented quantum mechanically.
Controversy however remains about if this is ``the right''
implementation of a Hamiltonian constraint. For instance, it was
noticed that a similar implementation in a space of non-diffeomorphism
invariant states also yielded an Abelian algebra \cite{LeMa}. The
constraint also appears to contain a rather large number of spurious
solutions. For instance, applying the Thiemann construction in $2+1$
dimensions \cite{Th}, one encounters many quantum states in addition
to the usual solutions of the Witten quantization. In this case one
can remove the undesired states by the choice of inner product, and
this construction works rather naturally in $2+1$ dimensions. In $3+1$
dimensions, an example of potentially spurious solution is to consider
states $<\psi|$ with support on spin networks with regular
(non-extraordinary) vertices.  Since the action of Thiemann's
Hamiltonian on a bra state is to remove an extraordinary line (a line
ending in two vertices that are planar and with two of the three
incoming lines collinear), a state $<\psi|$ that does not contain
extraordinary lines is automatically annihilated by the
constraint. These states are quite problematic since it is difficult
to imagine how a semiclassical theory could be built on them that did
not approximate an arbitrary metric, including metrics that do not
satisfy the Einstein equations. Getting rid of undesired quantum
states is tantamount to ``imposing the Einstein equations'', and
therefore is expected to be a difficult task in $3+1$ dimensions. It 
s therefore not entirely surprising that it was possible to do it
in $2+1$ dimensions. These concerns are in our view enough to motivate
an active program of searching for alternatives to Thiemann's
quantization, although as should be evident from the above discussion,
{\em do  not imply }that there is something definitely ``wrong''
about the construction up to now. It might be that in the end
Thiemann's quantization {\em does} lead to the correct theory of
quantum gravity, albeit via an elaborate choice of inner product.

The aforementioned mathematical techniques have also had an impact in
the construction of path integrals for general relativity, an approach
that has come to be known as ``spin foams'' (see \cite{Baez} for a
recent review and references). Initial interest in this approach arose
quite independently of gravity, in the study of topological field
theories. In the spin foam approach to topological field theories one
expands the partition function of the theory in terms of the basis of
gauge invariants constructed with spin networks and performs the
integral over connections of the path integral. To perform this
integral one goes to the dual lattice, and is left with an expression
that is function of the valences associated to the faces of the dual
lattice. One can understand the resulting path integral as a time
evolution. If one slices the ``spin foam'', the intersections of the
faces of the dual lattice with a plane produce lines associated with a
spin inherited from that of the face of the dual lattice, 
that is one reconstructs a ``spatial'' spin network. When one expands the
action, one chooses a discretization of the expression. To recover the
continuum theory one therefore has to either refine the discretization
indefinitely or perhaps perform a sum over all possible
discretizations in the hope that the sum will be dominated by the
finer discretizations. In general these procedures produce
difficulties. Refining the lattice is problematic to implement in
practice with irregular lattices \cite{zapata}, and for a
non-renormalizable theory is very likely to lead to divergences even
if one were to use regular lattices (which in addition may conflict
with diffeomorphism invariance) to perform the refinement. The general
attitude has therefore favored the idea of summing over all
triangulations as a way to handle this issue.  In the case of
topological field theories, since they only have a finite number of
degrees of freedom, the resulting expression for the discretized
action happens to be invariant under choice of discretization. This
immediately simplifies things, since one does not need to sum over
triangulations, and accounts in part for the success achieved by this
approach in topological field theories. One immediately is left with a
discretized partition function that correctly embodies the dynamics of
the theory in a consistent way. This ``miracle'' is unlikely to repeat
itself for theories with an infinite number of degrees of freedom like
general relativity. Although the theory is invariant under
diffeomorphisms, it is by no means expected that the partition
function should be invariant under changes of triangulations not
related by diffeomorphisms. Worse, if one attempts to simply ``sum
over all triangulations'' one is clearly summing (infinitely) many
times the same triangulation shifted by four-dimensional
diffeomorphisms. The result will be divergent. It is akin to the
observation by Mazur and Mottola \cite{MaMo} in the context of
traditional path integrals for gravity that the resulting partition
function is divergent if one does not properly gauge fix the
theory. Unfortunately, it appears difficult that one will be able to
properly gauge fix in terms of spin networks, or alternatively, it
appears as difficult as handling the Hamiltonian and diffeomorphism
constraints. The situation appears particularly complex, since
space-time diffeomorphisms are implemented in a non-trivial way. For
instance, one can consider a pair of initial and final spin network
states $|s_i>$ and $|s_f>$ and {\em many} spin foams that interpolate
between them. Several of these {\em topologically different} (not
related by diffeomorphisms) spin foams may correspond to the same
space-time interpolating between the initial and final states. So it
is not just a matter of simply considering ``floating lattices'' to
get rid of the redundancy in the sum implied by diffeomorphism
invariance. An extreme example of this point is given by $BF$
theories, where {\em all} spin foams interpolating between $|s_i>$ and
$|s_f>$ yield the same result, no matter if they are related by
diffeomorphisms or not.  Furthermore, the sum over all spins involved
in the discretization of the action has also proved to be divergent in
several cases, although this divergence can be seen as an ``infrared''
problem and can be handled by the introduction of a cut-off. The
evolution operators depend on the cutoff but only through a fixed
overall factor. In fact surprisingly encouraging recent results have
been reported in regularizing the sums (for a given discretization),
even for the case of the Lorentzian path integral (see \cite{CrPeRo}
and references therein).

It is worthwhile noticing that if one were able to complete a
spin foam quantization of general relativity, one could also use this
to answer some of the issues arising from the Hamiltonian
approach. The spin foam approach allows to construct evolution
operators and therefore to construct functions of spin networks that
should be annihilated by the Hamiltonian constraint. The evolution
operators also embody in a finite way the infinitesimal symmetries
implied by the constraints of the canonical theory, so they naturally
lead to insights about the nature of the constraints. 

In this paper we would like to present the construction of a theory
inspired by spin foams which is associated with a Hamiltonian
constraint that can be explicitly solved. The theory we present is
quite remarkable in the light of the discussion above: it is well
defined in spite of the fact that it has an infinite number of degrees
of freedom. Moreover, we will see that analyzing the connection of the
evolution operators of the theory with the Hamiltonian picture one
discovers that the theory produces solutions to the Hamiltonian
constraint proposed by Thiemann. This makes the theory a candidate for
quantum theory of gravity.

The theory we will propose is derived from the ``spin foam''
formulation of BF theory. The latter is a topological field theory (in
either three or four space-time dimensions) whose solution space
corresponds to flat connections. Ooguri has proposed a partition
function for these theories \cite{ooguriMPL}, in terms of which one
can construct evolution operators (since these are totally constrained
theories, these operators are also projectors, in the sense that
acting on an arbitrary ``initial'' state they produce a solution of
the quantum Hamiltonian constraint of the theory). Since the evolution
operators are projectors, evolved states are left invariant by further
evolution. Such quantum states of BF theory are functions of spin
networks that also happen to solve the Hamiltonian constraint of
quantum gravity. This is not hard to believe, since flat connections
solve the constraints of quantum gravity and these states are
associated with flat connections.

The condition for the states to be kept invariant under evolution is
encoded in the following elementary ``moves'' in terms of which all
evolutions from one spin network to another can be achieved,
\begin{equation}
\psi
\left(
\raisebox{-10mm}{\psfig{file=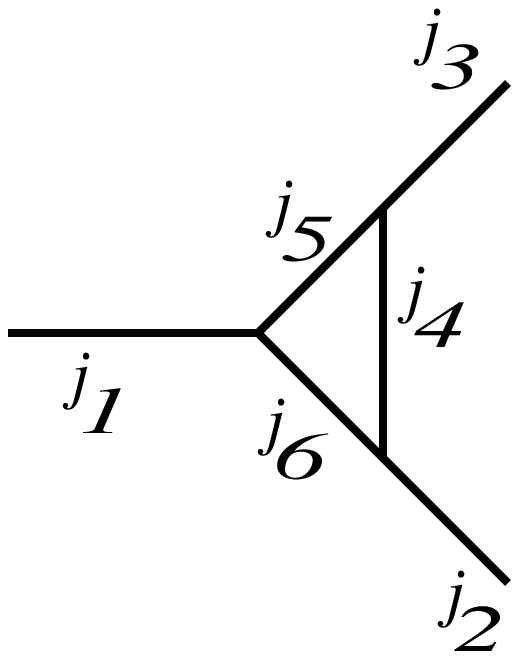,height=20mm}}\right) =
(-1)^{\sum_{i=1}^6 j_i}
\Lambda^{-1/2} \prod_{i=4}^6 \sqrt{2j_i+1} 
\wsymbol j_1/j_2/j_3/j_4/j_5/j_6/ \psi
\left(
\raisebox{-10mm}{\psfig{file=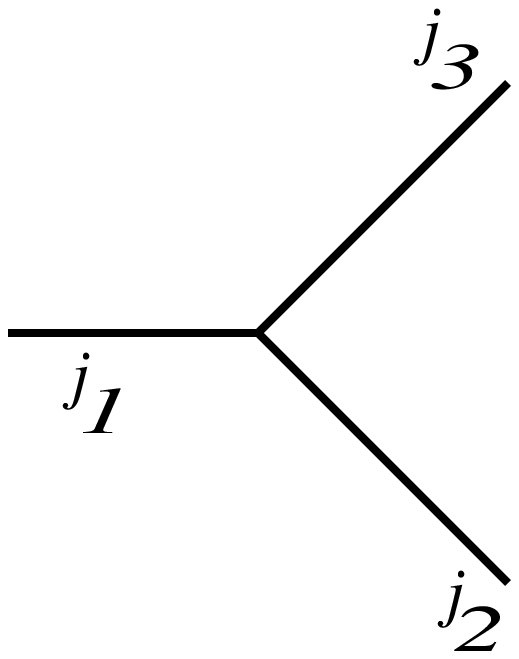,height=20mm}}\right),\label{u31}
\end{equation}
\begin{equation}
\psi
\left(
\raisebox{-10mm}{\psfig{file=ver.eps,height=20mm}}\right) =
\sum_{j_4,j_5,j_6}
(-1)^{\sum_{i=1}^6 j_i}
\Lambda^{-1/2}\prod_{i=4}^6 
\sqrt{2j_i+1} 
\wsymbol j_1/j_2/j_3/j_4/j_5/j_6/ \psi
\left(
\raisebox{-10mm}{\psfig{file=tri.eps,height=20mm}}\right),\label{u22}
\end{equation}
\begin{equation}
\psi
\left(
\raisebox{-7.5mm}{\psfig{file=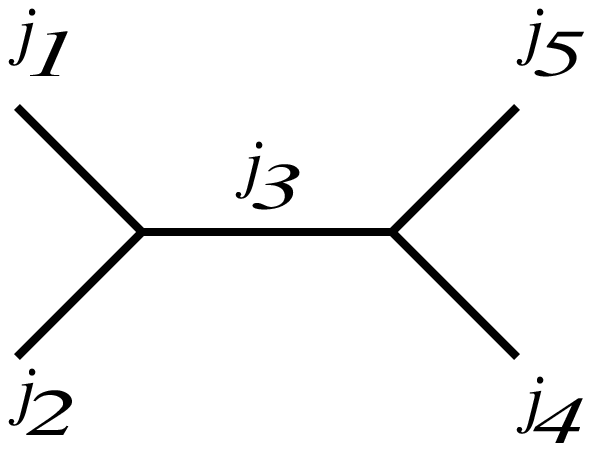,height=15mm}}\right) =
\sum_{j_6}
(-1)^{\sum_{i=1}^6 j_i}
\sqrt{2j_6+1}\sqrt{2j_3+1} 
\wsymbol j_1/j_2/j_3/j_4/j_5/j_6/ \psi
\left(
\raisebox{-7.5mm}{\psfig{file=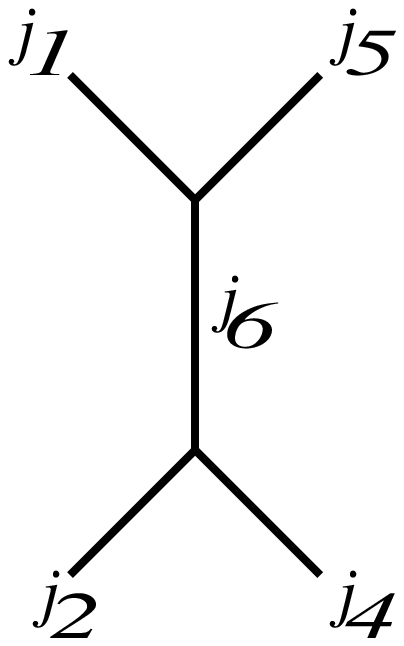,height=15mm}}\right).\label{u13}
\end{equation}
where $\Lambda$ is a cutoff that is needed for the Ooguri action to be
finite. The moves allow to untangle any spin network into a trivial
one, generating $6j$ symbols and other coefficients in the process.
These moves are well known, they are called recoupling identities and
just state that a spin network state is based on a flat
connection. For that reason all networks can be reduced to a trivial
one (the trivial network depends on the topology of the manifold, for
instance on a sphere it is a point, on a torus it is a ``Theta-net'').

The first move is interesting, since it action appears as the
``inverse'' of the action of Thiemann's Hamiltonian constraint for
general relativity. In Thiemann's construction the action of a
Hamiltonian constraint is to add a line at a trivalent
intersection. The first move allows to remove such a line. But in
fact, the result is stronger. It was shown that if one uses the first
move to ``undo'' the action of Thiemann's Hamiltonian constraint, the
end result vanishes \cite{GaGrPu}. That is, a quantum state whose
definition incorporates the first move automatically satisfies the
Hamiltonian constraint of quantum gravity (in fact it also solves the
generalization of the constraint proposed in \cite{GaRo} as well). We
will use this fact to construct the theory.

The theory we propose is defined in the following way: its
wavefunctions are defined by diffeomorphism invariant spin network
states that satisfy {\em the first two moves} of the three listed
above. These moves are inverse of each other so the theory is
consistent. It is well defined. But the lack of the third move
prevents us from ``undoing'' a non-trivial spin network into the
trivial one. The theory therefore has infinitely many inequivalent
states, which hints to the fact that in its connection representation
version the wavefunctions are not concentrated on flat connections
anymore and the theory has an infinite number of degrees of
freedom. Yet, due to the discussion of the previous paragraph, {\em
its wavefunctions still solve the Hamiltonian constraint of quantum
gravity as proposed by Thiemann}. This is the main result of this
paper. We have just constructed, simply by removing the last move, a
theory that is finite, well defined and whose states are in the kernel
of Thiemann's Hamiltonian constraint. The theory is well defined in
the sense that the evolution operators that arise from the above moves
are finite and well defined operators and they satisfy the condition
of being projectors, $\sum_{s'} P(s,s')P(s',s'')=P(s,s'')$, that as we
discussed above was required of evolution operators of a totally
constrained theory. The sum over the intermediate spin networks $s'$
is appropriately restricted (otherwise there is potential for a
divergence). In BF theory, as Ooguri
\cite{ooguriMPL} first discussed, the sum is only over colorations of
a given triangulation.  Choosing different triangulations just
correspond to different representations of the same Hilbert space. In
the case of our theory one needs a more subtle structure. The sum is
over all colorations and over all inequivalent ``skeletons'' of spin
networks. A skeleton is defined as the minimal spin network one
obtains when all triangles are removed. Since the projectors  are
nonvanishing only if the initial and final spin network share the same
skeletonization (the moves keep the skeleton invariant), then the left
hand side is indeed finite even if one sums over all
skeletons. Different spin networks with the same skeleton correspond,
as in Ooguri's case, to different representations of the Hilbert
space.

The construction works in three
and four dimensions and is not confined to trivalent intersections. If
one wishes to consider intersections of higher valence (which is
especially of interest in four dimensions) one need to consider
additional recoupling moves. For four valent intersections the move to
consider is \cite{ooguri4},
\begin{equation}
\Psi\left(
\raisebox{-15mm}{\psfig{file=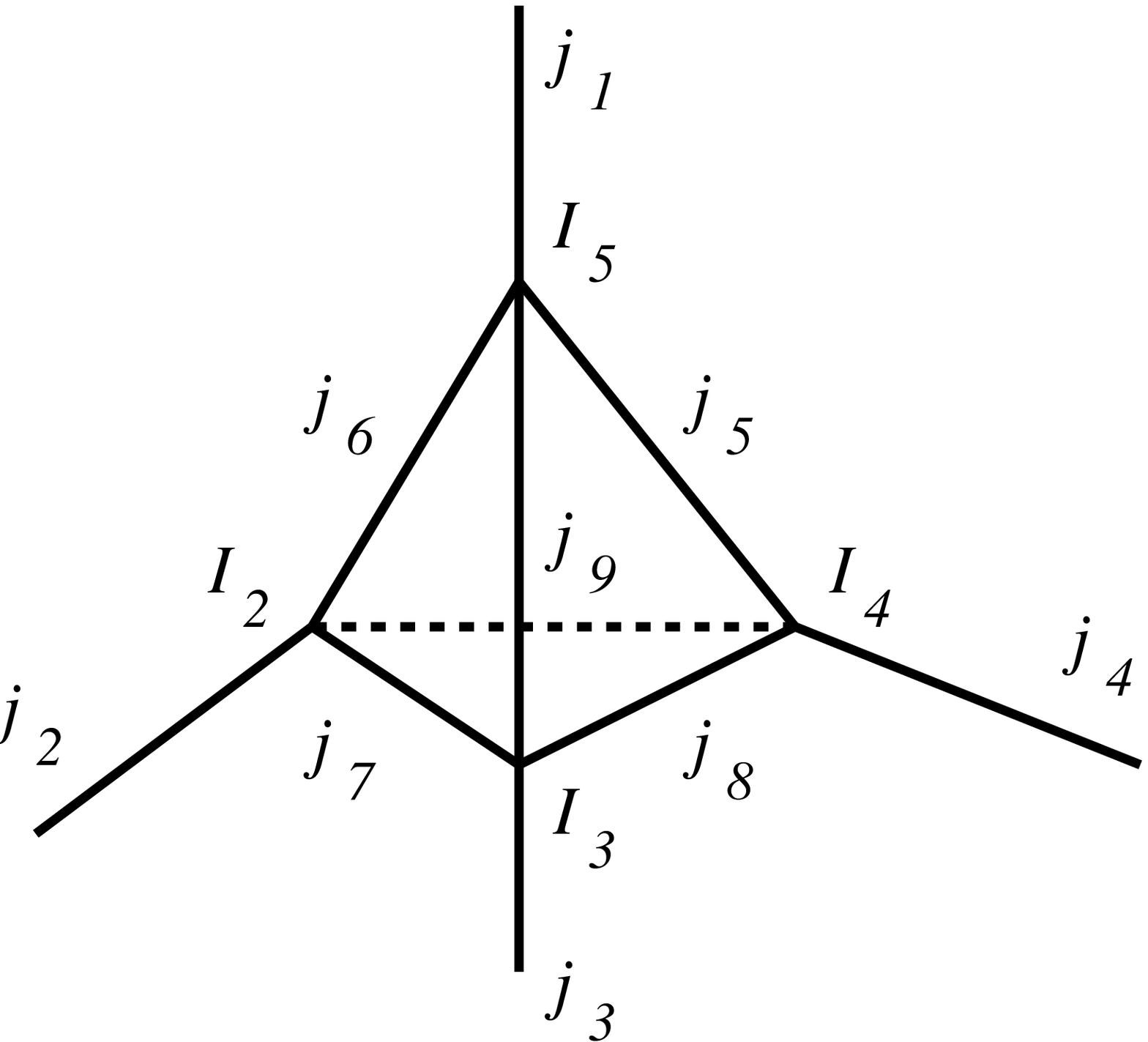,height=30mm}}\right)=
\prod_{i=5,k=2}^{10,5}\sqrt{2j_i+1}\sqrt{2 I_k+1} \left\{6j\right\}
\left\{15j\right\} 
\Psi\left(
\raisebox{-12mm}{\psfig{file=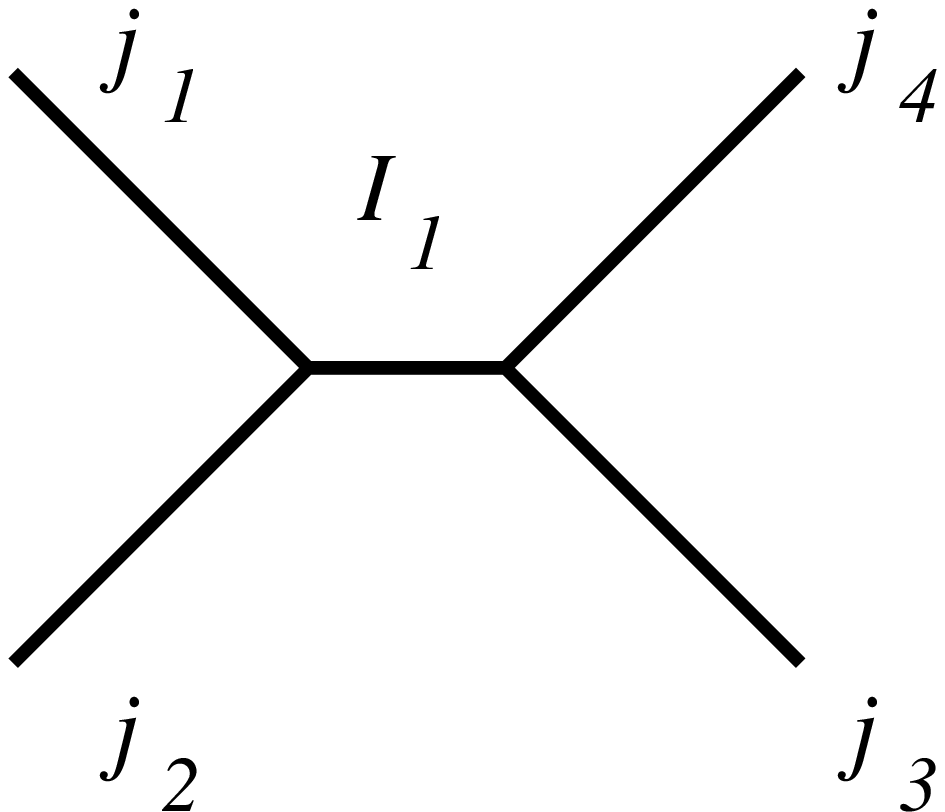,height=24mm}}\right)
\label{quedar}
\end{equation}
and its inverse. These moves contain the ones listed before as
particular cases by setting valences to zero.  One would need two
additional moves in order to have complete recoupling corresponding to
a flat connection if one has four valent intersections.

We need to elaborate a bit on the precise nature of the space of
states that we are proposing. In spite of similarities with gravity,
BF theory has an evolution operator that acts on a space of {\em
combinatorial} spin networks. The evolution operator ``propagates in
space'' in the sense that its action creates new vertices.  These
aspects seems to imply that there should be little connection between
these evolution operators and Thiemann's Hamiltonian constraint. The
latter operates on regular (not combinatorial), diffeomorphism
invariant, spin networks. Its action is only concentrated at vertices
and only produces new ``extraordinary'' vertices that are not ``seen''
if one further operates with the constraint. Yet, it is remarkable
that a state that is left invariant by our evolution operator (
acting on the space of diffeomorphism invariant spin networks) manages
to be annihilated by the Hamiltonian constraint of Thiemann. It can be
seen as if the condition implied by our theory is ``stronger'' on the
states than the one implied by the vanishing of Thiemann's
Hamiltonian. To give an analogy (it has only a partial meaning as we
will soon discuss), consider the Hamiltonian constraint of classical
general relativity (indices omitted) EEF=0. One can consider a theory
whose constraint is F=0 and therefore all its solutions will also be
annihilated by EEF=0.  The analogy here would be BF theory. It has
been known for some time \cite{GaGrPu} that the solutions of BF theory
(chromatic evaluations) are trivially annihilated by Thiemann's
Hamiltonian. The theory we are proposing today would be roughly of the
same kind as the one defined by a ``Hamiltonian constraint'' EF=0. All
its solutions would still be included in Thiemann's theory, but it has
a richer solution space than that of the theory defined by F=0. In
reality this analogy is too naive. The realization of the Hamiltonian
constraint of our theory in terms of classical variables is unknown,
but given the way we constructed it, is very likely (as we mentioned
before) to be highly non-local (it is ``simple in knot space'', which
suggests a very complex nature in connection space).

We have therefore constructed a well defined theory, with an infinite
number of degrees of freedom, which manages to solve in the sense
discussed above, the Hamiltonian constraint of (Euclidean) quantum
gravity as proposed by Thiemann. The theory exists in either three or
four dimensions. In three dimensions it is of little physical
interest, since there gravity is finite dimensional and
therefore the theory we constructed is clearly unphysical. The four
dimensional theory we constructed, since its states manage to solve
Thiemann's Hamiltonian constraint, is in principle a candidate for a
quantum theory of (Euclidean) gravity in four dimensions. It should be
evident from the way we constructed the theory (it was not derived
from an action, it is only formulated in terms of moves by removing
one of the moves of BF theory) that it is unlikely that it will be
connected with the correct physics of four dimensional quantum
gravity. Nevertheless, we believe it is a valuable example in that it
embodies many features considered desirable in a theory of quantum
gravity.  It is consistent in the sense that the evolution operators
are projectors. There are
explicit solutions to the Wheeler-DeWitt equation.  The evolution
operators do not suffer the ``locality''
\cite{Sm} issues that apparently arise in Thiemann's formulation. We
expect that when further progress is achieved in the analysis of a
semi-classical limit for spin-network based theories
\cite{semi}, our theory could be analyzed and ruled out as not
containing the correct classical physics of general relativity.  Both
the theory in three and the one in four space-time dimensions will
however be quite non-trivial and rich examples to be analyzed, which
go beyond the flatness of BF theory and nevertheless are finite and
well defined. The fact that the states solve the Hamiltonian
constraint of quantum gravity and include ``propagation in space''
sheds further light on the Hamiltonian proposed by Thiemann and
may even imply that the theory constructed has more physical
relevance than the one we can establish today.

We wish to thank Abhay Ashtekar for discussions.  This work was
supported in part by grants NSF-PHY0090091, NSF-PHY-9800973,
NSF-INT-9811610, by funds of the Horace C. Hearne Jr. Institute for
Theoretical Physics, and the Uruguay Fulbright commission.

\end{document}